%% file: Technical_Report.tex
\documentclass[sigconf]{acmart}

\AtBeginDocument{%
\providecommand\BibTeX{{%
\normalfont B\kern-0.5em{\scshape i\kern-0.25em b}\kern-0.8em\TeX}}}
\input{tools}

\setcopyright{acmcopyright}
\copyrightyear{2022}
\acmYear{2022}
\acmDOI{XXXXXXX.XXXXXXX}

\acmConference[WSDM'22]{WSDM conference}{February, 2022}{Tempe, Az, USA}
\acmPrice{15.00}
\acmISBN{978-1-4503-XXXX-X/18/06}

\begin{document}

\title{HTGN-BTW: Heterogeneous Temporal Graph Network with Bi-Time-Window Training Strategy for Temporal Link Prediction}

\author{Chongjian Yue}
\authornote{Work performed during the internship at MSRA}
\affiliation{%
\institution{Northeastern University}
\city{Shenyang}
\country{China}}
\email{20184545@stu.neu.edu.cn}

\author{Lun Du}
\authornote{Corresponding Author}
\affiliation{%
\institution{Microsoft Research Asia}
\city{Beijing}
\country{China}}
\email{lun.du@microsoft.com}

\author{Qiang Fu}
\affiliation{%
\institution{Microsoft Research Asia}
\city{Beijing}
\country{China}}
\email{qifu@microsoft.com}

\author{Wendong Bi}
\authornotemark[1]
\affiliation{%
\institution{Institute of Computing Technology, Chinese Academy of Sciences}
\city{Beijing}
\country{China}}
\email{biwendong20@mails.ucas.ac.cn}

\author{Hengyu Liu}
\authornotemark[1]
\author{Yu Gu}
\affiliation{%
\institution{Northeastern University}
\city{Shenyang}
\country{China}}
\email{1710589@stu.neu.edu.cn}
\email{guyu@mail.neu.edu.cn}

\author{Di Yao}
\authornote{Work performed during the visiting at MSRA}
\affiliation{%
\institution{Institute of Computing Technology, Chinese Academy of Sciences}
\city{Beijing}
\country{China}}
\email{yaodi@ict.ac.cn}

%\affiliation{%
%\institution{Northeastern University (China)}
%\city{Shenyang}
%\country{China}}
%\email{@mail.neu.edu.cn}

% \renewcommand{\shortauthors}{??? and >>>, et al.}

\begin{abstract}
With the development of temporal networks such as E-commerce networks and social networks, the issue of temporal link prediction has attracted increasing attention in recent years. The Temporal Link Prediction task of WSDM Cup 2022\footnote{https://www.dgl.ai/WSDM2022-Challenge/} expects a single model that can work well on two kinds of temporal graphs simultaneously, which have quite different characteristics and data properties,  to predict whether a link of a given type will occur between two given nodes within a given time span. Our team, named as \textit{nothing here}, regards this task as a link prediction task in heterogeneous temporal networks and 
proposes a generic model, i.e., Heterogeneous Temporal Graph Network (HTGN), to solve such temporal link prediction task with the unfixed time intervals and the diverse link types. 
That is, HTGN can adapt to the heterogeneity of links and the prediction with unfixed time intervals within an arbitrary given time period. 
To train the model, we design Bi-Time-Window training strategy (BTW)
which has two kinds of mini-batches from two kinds of time windows. As a result, for the final test, we achieved an AUC of 0.662482 on dataset A, an AUC of 0.906923 on dataset B, and won 2nd place with an Average T-scores of 0.628942. 
\end{abstract}

\keywords{link prediction, temporal network, heterogeneity, graph neural network}

\maketitle
\begin{table*}
\caption{Statistics of Dataset}
\label{tab:dataset}
\begin{tabular}{ccccc}
\toprule
Dataset&Node Number&Link Number&Relation Number&Prediction Period Size\\
\midrule
A&19,442&27,045,268&248&About 1200h\\
B&810,255&8,278,431&14&About 2300h\\
\bottomrule
\end{tabular}
\end{table*}

\section{Introduction}
Graph, as a general data structure to represent relationships between entities, is ubiquitous in the real world. Many of them are temporal graphs that are dynamic and evolving over time \cite{du2018dynamic,zhou2018dynamic}, such as E-commerce networks, social networks and communication networks \cite{wang2019tag2vec,wang2019tag2gauss}.
In such real graphs, one popular demand is temporal link prediction which is to predict new links in the future according to the historical information, such as recommending products to users in E-commerce networks, recommending friends to users in social networks, etc \cite{divakaran2020temporal,song2020inferring}.

In the Temporal Link Prediction task of WSDM Cup 2022, the 
committee
provides two datasets, dataset A and B. Dataset A is a dynamic event graph, in which entities are represented by nodes and different types of events are represented by different kinds of links.
Both nodes and links have related features.
Dataset B is a user-item graph in which users and items are represented by nodes and various types of interactions are represented by links. But only the links have
features.
For the task, the participants are required to predict whether a link of a given type will occur between two nodes within \textbf{a given time span}. The size of the prediction time span can vary arbitrarily within a certain range. And the time intervals between the training set and all prediction time spans are unfixed which makes the task is unusual compared with normal prediction tasks.

Many existing works have been proposed to tackle the temporal link prediction problem \cite{divakaran2020temporal,10.1007/s11280-017-0463-z,xu2020inductive,xue2020modeling,zhang2020multi,yin2019dhne,wang2020dynamic,yang2020dynamic,luo2020dynamic,fan2021heterogeneous,ji2020temporal,chen2022inferring}. TGN \cite{rossi2020temporal,ji2021dynamic}, as a representative work, presents a general deep learning framework for this task based on memory mechanisms and graph-based operators. However, compared to our problem in the challenge, previous works mainly consider a more simple scenario where the temporal graph datasets are homogeneous and without feature missing problems. 
Besides, few works involve the prediction of unfixed time spans which means that both the size of the prediction span and the time interval between the training set and the prediction are unfixed. In summary, the challenges we faced include (1) heterogeneity of graph data, (2) node/link feature missing problem, and (3) unfixed time-span prediction.

In this paper, we propose a new model Heterogeneous Temporal Graph Network (HTGN) with a new Bi-Time-Window (BTW) training strategy to tackle the above challenges. Similar to TGN, our model also has a memory mechanism and a neighborhood information aggregation module but with different updating strategies, and several new modules. 
Specifically, to tackle the first challenge, we introduce a relation embedding module to represent the type of links, and we have different prediction modules for each type of link to output an occurrence probability for such type of link. For the second challenge, we use a new feature-independent node embedding module to replace the raw node features to avoid the feature missing problem. For the third challenge, we introduce new prediction algorithm in the inference stage and BTW strategy in the training stage. To be specific, we split a given time span to several small time slices in the inference stage, and each small time slice can be viewed as a time point. There, our model can still conduct time point prediction, and then use an aggregation algorithm method to obtain the final prediction for the whole time span. In the training stage, we adopt two time windows, a memory window and a testing window, and training samples within the memory window are only used to update the memory module while the samples within the testing window are only used to update the model parameters. The BTW strategy can simulate the scenario of the final evaluation. Besides, we adopt a time interval encoder to deal with unfixed time span problem.

\section{Dataset}
The committee gives two datasets, dataset A and B, which have many differences. And each dataset contains a training set, an intermediate test set, and a final test set. For the intermediate test set, we can get the evaluation result in time. We will briefly introduce both datasets and our task as clear as possible.

\subsection{Dataset Overview}
Dataset A is a dynamic event graph, in which entities are represented by nodes and different types of events are represented by links. Both nodes and links have related features. Dataset B is a user-item graph, in which users and items are represented by nodes and various types of interactions are represented by links. But only the links have features. Therefore, unlike dataset A, dataset B is bipartite.

For simplicity, we regard both datasets as dynamic heterogeneous information networks. In such networks, node represents entities in Dataset A and users or items in Dataset B, link represents events in Dataset A and interactions in Dataset B, relation represents link's type, and timestamp in Unix Epoch indicates when the link occurs. All links are undirected. And we ignore all raw features provided by datasets.

Formally, (\textit{source}, \textit{destination}, \textit{relation}, \textit{timestamp}) means that a link between source node and destination node of a given relation occurs at time of \textit{timestamp} in Unix Epoch.

Every timestamp can be divided by 3600 in the training set of A but can't in the dataset B and the test sets of A. For consistency, we convert all timestamps in both datasets in this way: 
$$
timestamp_{new}=\lfloor \frac{timestamp_{old}}{3600}\rfloor
$$
Then, we can convert the unit of all timestamp from seconds to hours.

\subsection{Task Definition}
We define the task as a temporal link prediction: Given link data before time \textit{T}, we predict whether a link of a given type will occur between two nodes within a given time span which is from time $T + 1$ to $T + x$ \cite{dunlavy2011temporal}. The time between $T + 1$ and $T + x$ is regarded as the prediction span of this prediction, and the period which contains all prediction spans is regarded as the prediction period of the dataset.

 Formally, the prediction task from test sets is the probability that (\textit{source}, \textit{destination}, \textit{relation}, $timestamp_1$, $timestamp_2$) is true. The time between $timestamp_1$ and $timestamp_2$ is its prediction span we defined. The time between the min $timestamp_1$ and the max $timestamp_2$ from all predictions is the prediction period of the dataset.
 
 For simplicity, we divide a prediction span into time points by hour. So we only need to predict the probability that (\textit{source}, \textit{destination}, \textit{relation}, \textit{timestamp}) is true, where the \textit{timestamp} is in hours. For these probability values from the same prediction span, we try many aggregation methods. Then we take the maximum as the corresponding result. Now, the data from the training sets and the test sets are same in form: (\textit{source}, \textit{destination}, \textit{relation}, \textit{timestamp}).

According to the words defined above, Table~\ref{tab:dataset} shows the statistics of dataset A and dataset B.
\begin{figure*}[h]
\centering
\includegraphics[width=\linewidth]{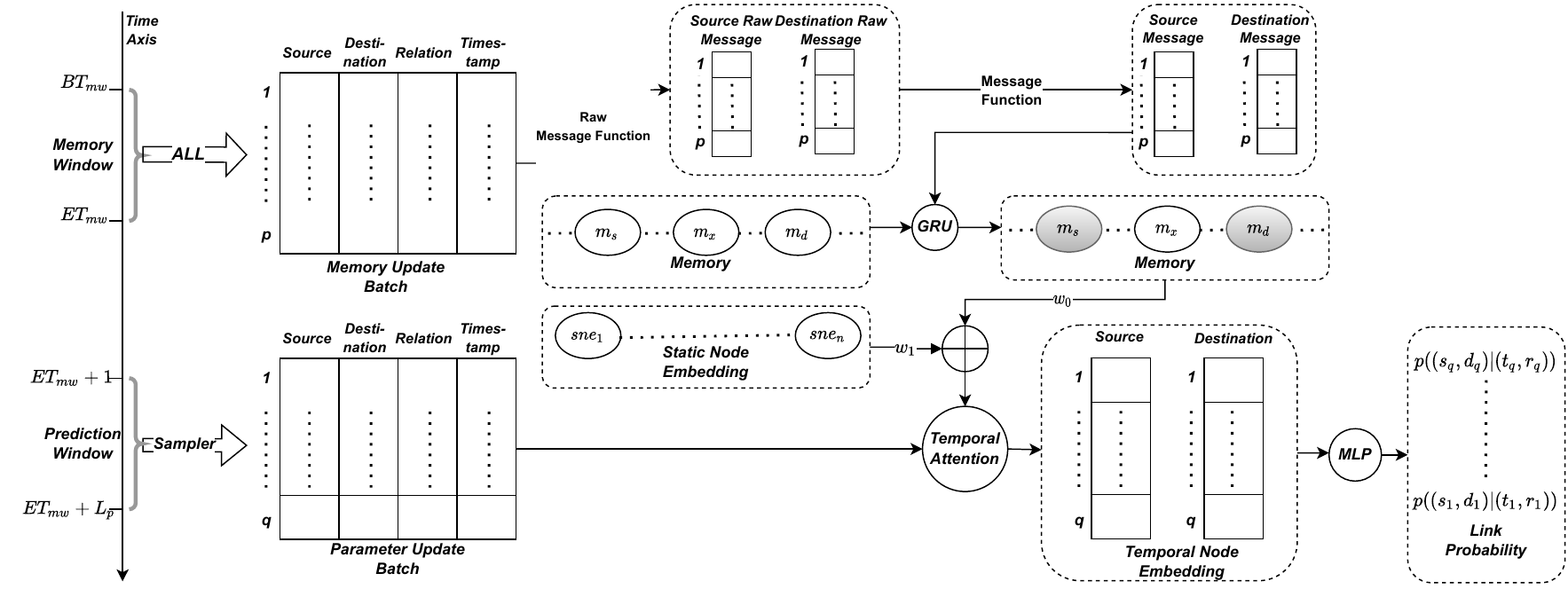}
\caption{Operation Flow of HTGN. $BT_{mw}$ is the beginning time of the memory window. $ET_{mw}$ is the end time of the memory window. $ET_{mw}+1$ is the beginning time of the prediction window. The length of the prediction window is $L_p$ hours. So the end time of the prediction window is $ET_{mw}+L_p$. }
\label{fig:model}
\Description{This is a description.}
\end{figure*}

\section{Heterogeneous Temporal Graph Network}
Heterogeneous Temporal Graph Network (HTGN) can be regarded as an encoder-decoder pair. The encoder, based on TGNs, is used to map from a temporal graph to the embeddings of the nodes $Z(t)=(z_1(t),z_2(t),...,$ $z_n(t))$ for each time $t$ \cite{rossi2020temporal}. And the decoder takes two embeddings as inputs and makes a link prediction. Figure \ref{fig:model} shows the structure of HTGN. Partial modules have been modified to meet our needs, but there are still some modules that are similar or identical to those in TGNs. We will introduce the modules of HTGN one by one.
\subsection{Modules}
{\bfseries Memory.}
Same as that in TGNs, the memory module consists of vectors $m_i(t_i)$ for each node $i$. $t_i$ is the memory update time of node $i$, and each node has its own memory update time. The memory of a node is initialized to a zero vector and updated when one link associated with this node appears. Note that the memory is not updated in backpropagation. This module is used to store nodes' historical information. More details can be found in TGNs. 

{\bfseries Static Node Embedding.}
It is a new module used to represent nodes' features. The static node embedding module also consists of vectors $sne_i$ for each node $i$, which are initialized randomly. Unlike the memory module, the static node embedding will be updated in backpropagation. It can be viewed as a substitute for raw node features so that we don't have to care whether a dataset has node features.

{\bfseries Relation Embedding.}
For each relation $r$, there is an embedding vector $re_r$ which is called relation embedding. By this module, we can ignore whether a dataset has raw relation features and improve the ability of HTGN to express the heterogeneity of graphs.

{\bfseries Time Encoder.}
It is an encoder that can map a time difference $t_2-t_1$ to a vector $te_{t_2-t_1}$. But different from TGNs, we use a time encoder with practical significance. 
% Problem!
Specifically, for each time difference, we calculate some numbers: the number of hours in a day, the number of days in a week, the number of days in a month, the number of weeks in a month, and the number of months in a year.
Then we concatenate these numbers to a vector, and encode this vector by a two-layer MLP which is activated by ReLU after each layer. There are three kinds of time differences: the link time minus the memory update time, the query time minus the link time, and the query time minus the memory update time. Correspondingly, we use three such time encoders in HTGN.

{\bfseries Raw Message Function.}
We take the message function in TGNs as the raw message function. For each link ($source$, $destination$, $relation$, $timestamp$), we will record two raw messages for updating the memory of the source node and destination node. A raw message is a simple concatenation of the source node memory $m_s$, the destination node memory $m_d$, the relation embedding $re_r$ and the time difference embedding $te_{t_l-t_s}$ for the difference between the link time $t_l$ and the source node's memory update time $t_s$:
\begin{align}
&rmsg_s(t_l)=m_s\parallel m_d \parallel re_r \parallel te_{t_l-t_s} \\
&rmsg_d(t_l)=m_d\parallel m_s \parallel re_r \parallel te_{t_l-t_d}
\end{align}
{\bfseries Message Function.}
The message function is a two-layer fully connected feed-forward network. The first layer is activated by ReLU, and the second layer is not activated but the output is batch normalized. This function is used to map a raw message $rmsg_i(t_l)$ of node $i$ to a message $msg_i(t_l)$.

After a batch of links, a node will have many messages. In TGNs, there is a message aggregator to aggregate these messages into one message. But we remove this module to better save temporal information in HTGN, which also brings an expensive time cost. 

{\bfseries Memory Updater.}
We use a GRU as the memory updater. It takes a node memory as the initial hidden state, messages of the corresponding node as the features of the input sequence, and the updated hidden state as the updated memory. Meanwhile, we take the max timestamp of these messages as the new memory update time.

{\bfseries Temporal Graph Attention.}
Similar to TGNs, this module is used to generate the temporal embedding $z_i(t)$ of node $i$ at any time $t$. We choose temporal graph attention (attn) as the basic implementation and make some adjustments to it for better performance. Now, this module is a one-layer graph attention to compute $i$’s embedding by aggregating information from its partial one-hop neighbors in our model:
\begin{align}
&z_i(t)=MLP(q_i(t)\parallel {o_i(t)}),\\
&{o_i(t)}=MultiHeadAttention(q_i(t),K(t),V(t)),\\
&q_i(t)=\widetilde{z}_i\parallel \phi_{src}(t-t_i),\\
&K(t)=V(t)=C(t),\\
&C(t)=[\widetilde{z}_j\parallel \phi_{ngh}(t-t_j^{\prime})\parallel re_{ij},...,\widetilde{z}_k\parallel \phi_{ngh}(t-t_k^{\prime})\parallel re_{ik}].
\end{align}
Here, $t$ is a prediction time, $t_i$ is the memory update time of $i$, $z_i(t)$ is the temporal embedding of $i$ at time $t$, $\widetilde{z}_i(t_i)$ is the initial temporal embedding of node $i$ which is the weighted sum of the memory $m_i(t_i)$ and the static node embedding $sne_i$. And we select some links $\{e_{ij}(t_j^{\prime}),...,e_{ik}(t_k^{\prime})\}$ related to $i$, whose link times are closest to $t_i$. $\{re_{ij},...,re_{ik}\}$ are relation embeddings of these links, $\{t_{j}^{\prime},...,t_{k}^{\prime}\}$ are timestamps of these links, and $\{\widetilde{z}_j(t_j),...,\widetilde{z}_k(t_k)\}$ are initial temporal node embeddings of these links' destination nodes. We omit $t_x$ from ${\widetilde z}_x(t_x)$ in the equations. $\phi$ is a time encoder we mentioned above: $\phi_{src}$ is only used to encode the time difference between the prediction time and the memory update time, and $\phi_{ngh}$ is only used to encode the time difference between the prediction time and the link time. $\parallel$ is the concatenation operator, and $MultiHeadAttention$ is an operation proposed in \cite{vaswani2017attention}.

{\bfseries X-MLP.}
For each relation, we take an MLP as the corresponding decoder, which takes two temporal node embeddings as input and outputs a value as the probability of a link with a given relation. The number of MLPs is the same as the number of relations. For acceleration, we calculate all MLPs in parallel and name it XMLP.

% two windows: a memory window && a prediction window
\subsection{Bi-Time-Window Training Strategy}

In the training stage, we set two adjacent sliding time windows, i.e., Memory Window and Prediction Window, with length $L_m$ hours and $L_p$ hours for the training batch generation. For the $\tau$-th training step, we select samples (i.e., temporal links) within Memory Window and Prediction Window to construct a memory update batch $B_\tau^{m}$ and a parameter update batch $B_\tau^{p}$, respectively. The samples in the memory update batch are only used to update memory and do not participate in the back propagation. The samples in the parameter update batch are the opposite only serving parameters update.

Intuitively, such a training strategy is more aligned with the real inference stage where we need to conduct predictions for a time span in a relatively far future. Thus, our model cannot leverage the memory which is updated by real links close to the prediction time. In our training stage, the length of Prediction Window $L_p$  is set to be similar to the time length between the latest training data and the latest test data, and the samples within the Prediction Window cannot update the memory in the $\tau$-th training step, which can be viewed as a simulation of the prediction procedure. 

To be specific, when generating the memory update batch $B_\tau^{m}$, we fix the number of samples instead of fixing time lengths for performance issues. When generating the parameter update batch $B_\tau^{p}$, we conduct a negative sampling to generate negative samples. We design two sampling strategies: pure random and varying along a certain dimension. Pure random is to randomly generate negative sample quadruples, (\textit{source}, \textit{destination}, \textit{relation}, \textit{time}) within Prediction Window. Varying along a certain dimension means that we select a dimension from the \textit{source}, \textit{destination}, and \textit{relation} in turn, then replace the corresponding dimension of the positive samples with a random value. Negative samples from two strategies combined with positive samples form the parameter update batch. 

Finally, the temporal graph attention module uses the initial temporal embedding to calculate the temporal embedding of the corresponding nodes in the prediction batch. As a decoder, X-MLP will give the link probability by a pair of temporal node embeddings. Lastly, through the ground truth and probability, we calculate the loss value and update parameters of HTGN by back propagation.

\section{Experiment}
\begin{table}
\caption{Experimental Configuration and Result}
\vspace{-0.3cm}
\label{tab:exp}
\begin{tabular}{ccc}
\toprule
 & Dataset A & Dataset B \\
\midrule
Relation Dimension & 64& 8 \\
Other Dimensions& 64& 128 \\
Attention Heads& 2 & 4 \\
Initial $w_0$ & 0.1 & 1.0 \\
Initail $w_1$ & 0.9 & 0.0 \\
Prediction Period& 1300h & 2300h \\
Memory Update Batch Size & 1024& 512 \\
AUC& 0.663338& 0.906327\\
\bottomrule
\vspace{-0.5cm}
\end{tabular}
\end{table}
In our experiments, we use the whole training set for model training and take the result on the intermediate test set as the measurement. We select the best model from 10 epochs to generate the final prediction result for dataset A, and 15 epochs for dataset B.

For both datasets, the learning rate is 0.001; the dropout is 0.1; the size of a memory update batch is 1024; the size of a positive prediction batch is 1024; the size of a negative prediction batch that comes from pure random is 1024; the size of a negative prediction batch that comes from the strategy of varying along a certain dimension is 3072; and we select 40 neighbors in the temporal graph attention module. More hyperparameters and results are reported in Table \ref{tab:exp}. Here, other dimensions include the message dimension, the memory dimension, the time encoder dimension, and the temporal node embedding dimension. The initial $w_0$ is the initial given weight of the memory module, and the initial $w_1$ is the initial given weight of the static node embedding module. AUC means the AUC score on the intermediate test set.

\section{Conclusion}
In this paper, we introduce our proposed model HTGN with a Bi-Time-Window training strategy in details. As a result of the challenge, we won the second place on the leaderboard.

\bibliographystyle{ACM-Reference-Format}
\bibliography{sample-base}

\end{document}

%% file: tools.tex
\usepackage{color, xcolor}
\usepackage{balance}